# Experimental Stick-Slip Behaviour In Triaxial Test on Granular Matter

## F. Adjémian & P. Evesque

Lab MSSMat, UMR 8579 CNRS, Ecole Centrale Paris
92295 CHATENAY-MALABRY, France, e-mail: evesque@mssmat.ecp.fr

**Abstract:**

*This paper is concerned with the quasi-static rheology of packings of glass spheres of diameter d (d=0.2mm, 0.7mm or 3mm). Stick-slip behaviour is observed on small spheres, i.e d=0.2mm & 0.7mm ; one observes also in this case a weakening of the rheology as the rate of deformation increases, and the larger the rate the larger the weakening; this generates macroscopic instabilities and stick-slip. Statistics of stick-slip events have been determined, which show that the larger the sample the more regular (i.e. "periodic") the sick-slip, the faster the strain rate the less periodic the events. One concludes that this stick-slip is generated at the macroscopic level and comes from the macroscopic rhelogical law. However when sample is small, local fluctuations perturb the macroscopic events and trigger them erratically.*

**Pacs # :** 5.40 ; 45.70 ; 62.20 ; 83.70.Fn

___________________________________________________________________

In general, monotonic axial compression of granular matter, of sand or of clay exhibit smooth mechanical response, independent of the deformation rate [1]. However some cases are not smooth but exhibit stick-slip [2]. The problem in this case is to understand whether this stick-slip behaviour is induced by the macroscopic law or if it is generated from erratic phenomena on a microscopic scale. For instance, it was argued in [2] that stick-slip observed in packing of spheres are induced by local effect due to regular arrangement. Here we pursue the investigation of [2] and conclude differently because the larger the sample the less erratic the stick-slip. Also the mechanism which generates the instability is identified.

In this work, statistics of stick-slip and its domain of existence are determined as functions of the lateral pressure $\sigma'_3$ (see section 1 for definition), of the rate of deformation $d\varepsilon_1/dt$, of the sample size. It appears that stick-slip is not periodic and does not depend on the strain rate history under present experimental conditions. Furthermore, the distribution of "waiting time" ($\Delta\varepsilon_1$) between successive stress falls is found exponential when piston speed is fast, whereas it is Gaussian-shaped at slower strain rate and in large samples only. It is also worth noting that strain localisation is observed at the end of the tests, but does not affect the stick-slip statistics. An other general trend is that the samples with stick-slip exhibit both smoothening and weakening when the rate of deformation $d\varepsilon_1/dt$ is increased.

So, these results show (i) that stick-slip is not generated by the fact that the packing is regular, (ii) that the smaller the sample the more erratic the stick-slip mechanism and the larger the relative fluctuations, (iii) the larger the sample the more periodic the stick-slip, the larger the stress falls, but the less fluctuating their amplitudes. So, these results allow to conclude that the observed stick-slip is linked to a macroscopic mechanism, induced by the macroscopic law itself; the mechanism of instability is attributed to the weakening process which is observed when the strain rate is increased. However stick-slip is non periodic and its size non regular, even in large sample, due to some erratic fluctuations at the microscopic scale.

## 1. Experimental data and reference behaviour

The experimental set-up is composed of a vertical cylindrical sample (diameter D, height H) of granular material maintained under partial constant vacuum in the pressure range $u_v$ = 40-70 kPa ; the material is made of glass spheres of diameter d (d =0.2mm or 0.7mm or 3mm). The sample is submitted to a monotonic 1-D axial compression at constant strain rate $d\varepsilon_1/dt$. The outside pressure is the normal pressure $\sigma_3$ = 100 kPa. Let define $\sigma'_i=\sigma_i-u_v$ ; the aim of a triaxial test experiment is to determine the evolution of the vertical overload $q=4F/(\pi D^2)=\sigma_1-\sigma_3=\sigma'_1-\sigma'_3$ as a function of the sample strain ($\varepsilon_1$ = - $\delta H/H$, $\varepsilon_v=\delta v/v$), for a given stress path, for instance $\sigma'_3=c^{ste}$ here.

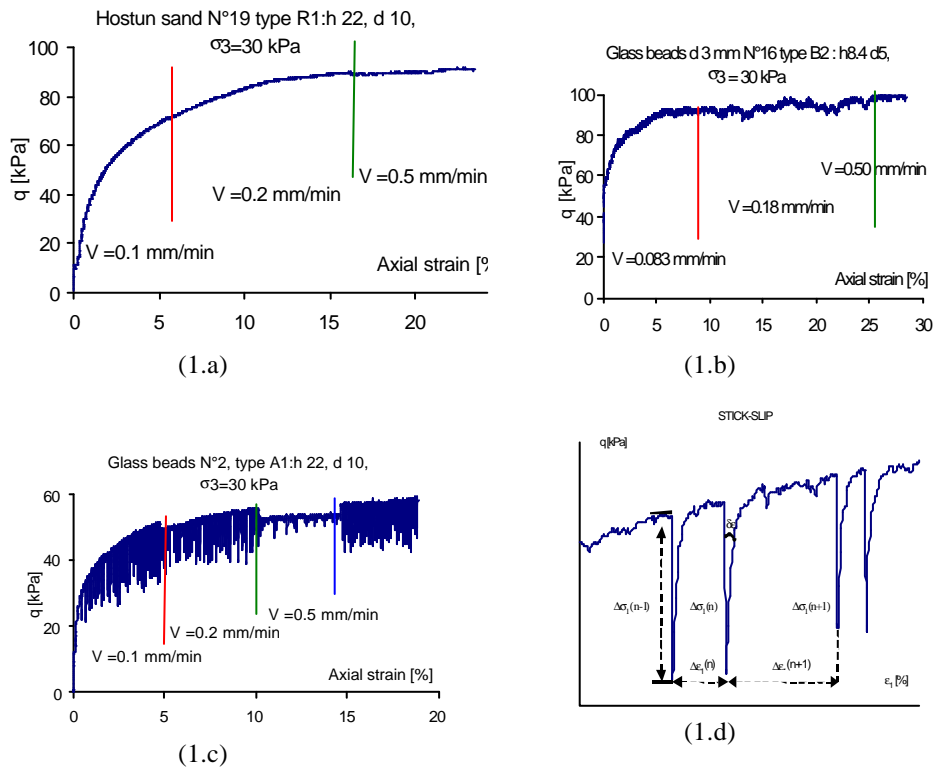

**Figure 1**: **The three different behaviours obtained from the different samples**. (a)- Classical Hostun sand behaviour (with H= 20 cm, D= 10 cm, $\sigma'_3$ =30 kPa). (b)- Typical fluctuations with d=3mm glass beads (with H= 8.4cm, D=5cm, $\sigma_3$ =30 kPa). (c)- Typical stick-slip behaviour on d=630-840 μm glass beads (with H= 22cm, D=10cm, $\sigma_3$=30 kPa). (d)- stick slip features: $\Delta q$ = amplitude of the abrupt deviatoric stress variation, $\Delta\varepsilon_1$= the strain between two stick-slips.

This paper wants to demonstrate that mechanical behaviour of a macroscopic ensemble of glass beads can be not "classic". Indeed, in the soil mechanics literature, a quasi-static regime is observed at slow strain rate $d\varepsilon_1/dt$, for which mechanical response does not depend on $d\varepsilon_1/dt$ under a monotonic 1-D vertical compression. These "classic" behaviours are well known [1]; we have observed them as usually on loose Hostun sand, *cf.* Fig. 1.a which reports the typical $q=4F/(\pi D^2)$ *vs.* $\varepsilon_1$ curve obtained at small $\sigma'_3$=30kPa. Also, in Fig. 1.a, the deformation rate $d\epsilon_1/dt$ has been changed twice during the experiment, at the vertical marks, but the curve looks and remains quite smooth and regular even when the strain rate $d\epsilon_1/dt$ is varied : that proves the existence of the quasi static regime.

Table 1 reports the characteristics of the studied samples. Fig. 2 characterises the spheres. This paper summarises the results of a series of 30 tests about, whose precise characteristics are reported in Table 2 at the end of the paper .

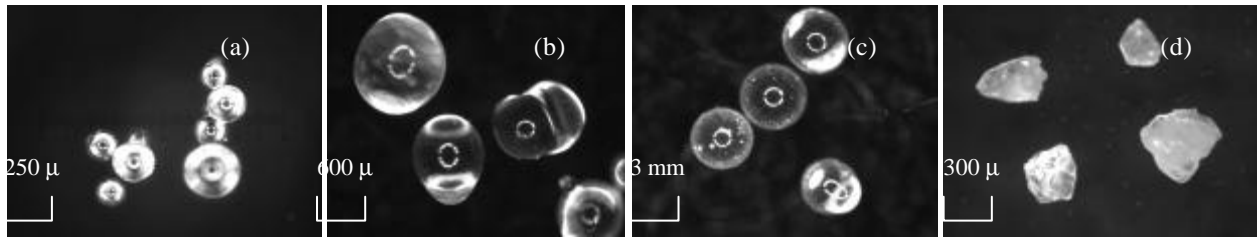

**Figure 2 : Photographs of the different glass beads** : (a) sample A', d= 150-250 µm from optical microscope (x100). (b) sample A, d= 630-840 µm from optical microscope (x50). (c) sample B, d= 3mm from optical microscope (x5). (d) Hostun sand from optical microscope: The geometry of the material surface is not spherical in this last case. (x50).

| type | H | D | d | Friction |
|---|---|---|---|---|
| **A1** | 221 mm | 101 mm | 630-840 µm | 31° ± 2.5° |
| **A2** | 84 mm | 51 mm | 630-840 µm | 37° ± 2.1° |
| **A3** | 50 mm | 50 mm | 630-840 µm | 41° ± 2.1° |
| **B2** | 84 mm | 51 mm | 3 mm | 37° ± 2.1° |
| **A'1** | 221 mm | 101 mm | 150-250 µm | 31° ± 2.5° |
| **R1** | 221 mm | 101 mm | Sand, d≈0.3µm | 31° ± 2.5° |

**Table 1: The characteristics of the 4 kinds of samples**: height H, diameter D, grain size d, and friction angle φ measured at large deformation.

## 2. Fluctuations and stick-slip behaviour

Fig. 1.b is a typical response obtained with large mono-disperse glass spheres (d=3mm): the $q=4F/(\pi D^2)$ *vs.* $\epsilon_1$ curve is rather noisy with a typical relative width $\delta q/q \approx 3\%$. It is likely due to the smallness of the sample size compared to the grain size ; indeed, curves obtained from computations with discrete element method [3-5] exhibit similar fluctuations amplitude.

Fig. 1.c reports a typical q *vs.* $\epsilon_1$ curve obtained with small poly-disperse glass spheres. Part of this curve has been expanded horizontally in Fig. 1.d, in order to exemplify the stick-slip mechanism and to define few quantities : the q *vs.* $\epsilon_1$ curve exhibits a series of rapid large stress diminution $\Delta q(n)$ at a discrete set of deformation $\epsilon_1(n)$ ; we call $\Delta\epsilon_1(n)= \epsilon_1(n)- \epsilon_1(n-1)$ the deformation separating two successive falls , $\delta\epsilon_1(n)$ the "duration" of fall n in strain unit and $q_{max}(\epsilon_1)$ the typical stress before each fall or after $\delta\epsilon_1$. Fig. 1.d shows that $\delta\epsilon_1 \ll \Delta\epsilon_1$ and that stress fall ($\Delta q$) is even much faster, since it occurs within a sampling time, *i.e.* 3s ; it corresponds then to negligible deformation. So, after this fall, deformation proceeds and q recovers its previous value

$q_{max}(\varepsilon_1)$ within a strain $\delta\varepsilon_1 \ll \Delta\varepsilon_1$. In Fig. 1.c, the strain rate $d\varepsilon_1/dt$ has been changed twice showing the smoothening of the stress-strain relation and the weakening of the material at large speed, *i.e.* the faster the $d\varepsilon_1/dt$, the smaller the $q_{max}(\varepsilon_1)$ and the smaller the number of stick-slips. Using a dynamical spring-block model it is easy to show that the weakening makes the dynamics unstable; so it generates stick-slip most likely.

## 3. Qualitative and quantitative results

We have observed that statistics on $\Delta q$, $\Delta\varepsilon_1$ and $q_{max}(\varepsilon_1)$, depend on the rate $d\varepsilon_1/dt$, but do not depend on the chronology of the applied strain rate. Furthermore, we have found that the slower the rate $d\varepsilon_1/dt$, the larger the fall $\Delta q$ and the more numerous the peaks. We have noticed also that the larger the sample, the less scattered the instabilities, for a given rate of deformation $d\varepsilon_1/dt$. We have found also that the relative average $<\Delta q>/q$ of stick-slips is identical for lateral effective stress $\sigma'_3$ equal to 60 kPa and equal to 30kPa, whereas the stick-slip behaviour disappears when the sample is dense.

Statistical analysis has been performed and typical examples reported in Figs. 3 and 4; it has demonstrated that the mean strain $<\Delta\varepsilon_1>$ between two stick-slips increases with the piston speed, whatever the sample size. It means that deep $\Delta q$ falls are rarer at larger speed; but the missing large events are replaced by smaller falls. This tends to show that stick-slip tends to disappear at fast strain rate [6].

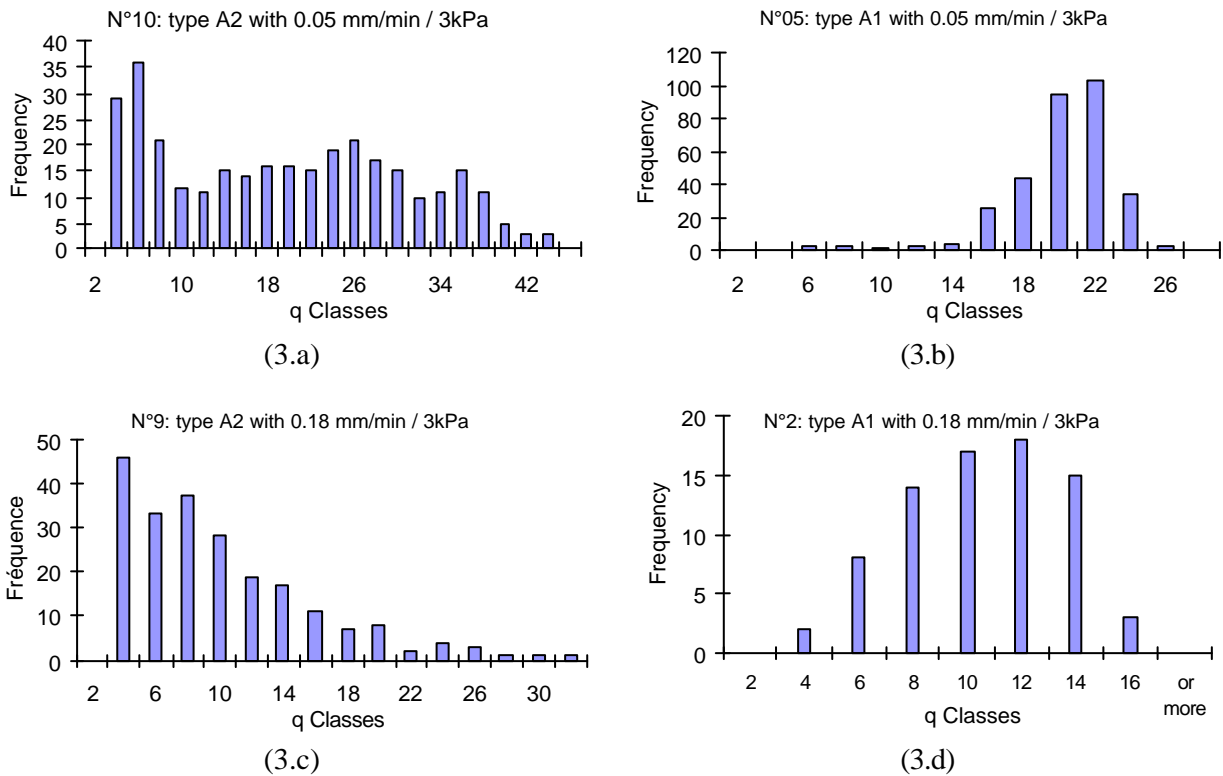

**Figure 3**: **Distribution of the $\Delta q$ size calculated with a minimum amplitude $\Delta q_{min} = 3$kPa**. The $\Delta q$ classes are expressed in kPa (a)-Test n° 10 type **A2**, with H= 8.4cm, D= 5cm, $\sigma'_3$ = 30 kPa, with V = 0.05 mm/min. (b)- Test n° 5 type **A1**, with H= 20cm, D= 10cm, $\sigma'_3$ = 30 kPa , with V = 0.05 mm/min. (c)- Test n° 9 type **A2**, with H= 8.4cm, D= 5cm, $\sigma'_3$ = 30 kPa, with V = 0.18 mm/min. (d)- Test n° 2 type **A1**, with H= 20cm, D= 10cm, $\sigma'_3$ = 30 kPa, with V = 0.18 mm/min

Distributions of waiting time $\Delta t_n = (\Delta\varepsilon_n \cdot H)/V$ have been determined. Two kinds of statistics are observed : the Gaussian distribution of $\Delta t_n$ reported on Fig. 4.a is observed only for large samples submitted to small piston speed (V<0.18mm/min). Whereas Fig. 4.b exhibits an exponential distribution for $\Delta t_n$ ; this last distribution is observed in small samples, or in large samples when piston speed is fast. So, owing to its exponential-like distribution, it may be considered that Fig. 4.b corresponds to the statistics of a single elementary events whose size fluctuates according to an exponential law ; within the same thought, each event of Fig. 4.a can be viewed as corresponding to the combination of few independent elementary events.

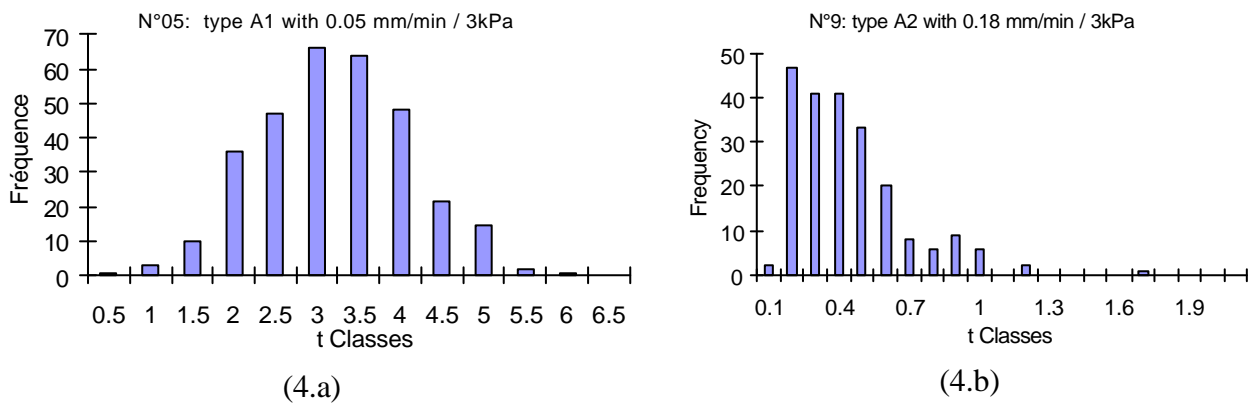

(4.a)    (4.b)

**Figure 4** : **Distribution of the $\Delta t$ size**. The $\Delta t$ classes are expressed in minutes. (a)- Test n° 5 type **A1**, with H= 20cm, D= 10cm, $\sigma'_3$ = 30 kPa , with V = 0.05 mm/min. (b)- Test n° 9 type **A2**, with H= 8.4cm, D= 5cm, $\sigma'_3$ = 30 kPa, with V = 0.18 mm/min.

Distributions of the $\Delta q$ falls were determined too; they are exponential for all tests with small samples, but become Gaussian too for large sample and slow rate, *cf.* Fig. 3.

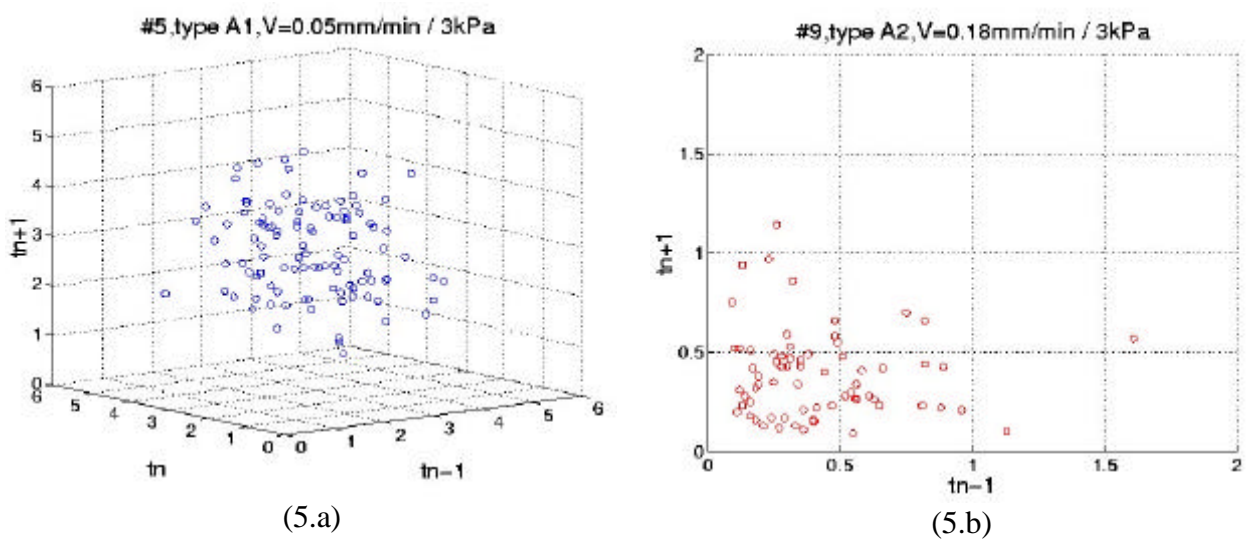

(5.a)    (5.b)

**Figure 5:** (a)-3D representation of test #5, large sample, with V=0.05 mm/min in the phase space ($\Delta t_{n-1}$, $\Delta t_n$, $\Delta t_{n+1}$). (b)- Projection in 2D of test #9, small sample, with V=0.18 mm/min in the plane ($\Delta t_{n-1}$, $\Delta t_{n+1}$).

To perform an analysis of the kind proposed by dynamical system theory, we tried to determine some attractor. Fig. 5 reports the trajectory of the stick-slip events, in a three-dimensional phase space $\{\Delta t_{n+1}, \Delta t_n, \Delta t_{n-1}\}$. For large samples submitted to very slow piston speed, a spherical cloud is formed, centred at $\{\Delta t_o, \Delta t_o, \Delta t_o\}$, *cf.* Fig. 5.a; this is compatible with the fact that each point corresponds to the sum of three independent, *i.e.* non correlated, events having each a Gaussian distribution of the kind $\exp\{(\Delta t-<\Delta t>)/\delta t\}^2$. Typical result on small sample is reported in Fig. 5.b ; a cloud is also observed, but its projection on any plane is rather triangular, which implies that each point is the sum of non correlated events with an exponential distribution of the kind $\exp\{\Delta t_n/\delta t\}$.

As the Gaussian distribution is found when the sample is large, and as its relative width decreases when the sample size gets larger, one shall extrapolate to larger samples and conclude that the relative width shall decrease to 0 in the macroscopic limit, and its mean $\Delta\varepsilon$ shall tend to a constant value. This stick-slip mechanics is then induced by the macroscopic rheological law and leads to periodic behaviour. However, when the sample is small enough, stick-slip is perturbed and triggered by local fluctuations ; this explain the exponential tail when the sample is small and the Gaussian distributions for larger (and finite) samples.

## 4. Conclusion

Stick-slip obtained during axial compression of a dry sample of glass beads is a complex phenomenon which depends on numerous parameters. Stress falls obtained with the smallest samples do not exhibit periodic behaviour, but present always a $\Delta q$ distribution which is exponential. Also, its "waiting time" distribution $\Delta\varepsilon/(\partial\varepsilon/\partial t)$ is often exponential too; this holds true except for large samples submitted to very small strain rate, which exhibit a Gaussian-like or a Poisson-like distribution. Stick-slip behaviour is severely disrupted for strain rate $\partial\varepsilon/\partial t$ above 0.50mm/min/H ;  then it tends to disappear. Both phenomena, *i.e.* stick-slip and its disruption, are likely related to the process of rheological weakening observed when increasing the strain rate. So, we believe that these effects are linked to the rheology at the macroscopic scale and not provoked by local fluctuations on microscopic samples. This analysis is strengthened by the analysis of the statistics as a function of the sample size: this statistics presents a cross-over from a single exponential to a Gaussian distribution; the single-exponential law is likely provoked by a single random event with a random size while the Gaussian distribution is generated likely by a complex event which is made of n independent elementary events of the previous kind .

Anyhow, it is obvious also from Fig. 1.c that the material is less "hard" at larger strain rate than at slower strain rate. This is able to provoke a dynamical instability at the macroscopic scale under some precise conditions. So, this instability generates probably the stick-slip mechanism at the macroscopic scale. If this hypothesis is true, the weakening shall also produce periodicity at macroscopic scale; however, this periodicity is broken in smaller samples by local fluctuations of stress and of stress-strain relation due to finite size effects.

Applied to seisms, which are macroscopic events occurring at a large length scale at the surface of the Earth, this study predicts that some rheological laws can generate large dynamics events; however, it shows also that these macroscopic events should be rather regular both in time and in size in homogeneous samples. So, it does not reproduce the scaling observed in nature. In the case of seism, this scaling is linked likely to the existence of complex structure at mesoscopic scale.

## Appendix: Table 2: The different experiments:

| | | | | | $\sigma'_3 = 30$ kPa | | | | | | | | | | | | | |
|---|---|---|---|---|---|---|---|---|---|---|---|---|---|---|---|---|---|---|
| Velocity | 0.05 mm/min | | | 0.083 mm/min | | | | 0.18 mm/min | | | | | 0.5 mm/min | | | | |
| size | H20, D10 | H8.4, D5 | H5, D5 | H20, D10 | H8.4, D5 | H5, D5 | H20, D10, d3 | H20, D10 | H8.4, D5 | H5, D5 | H8.4, D5, d3 | H20, D10 | H20, D10 | H8.4, D5 | H5, D5 | H20, D10 |
| # | #5 | #10 | #14 | #2 | #9 | #12 | #16 | #2 | #9 | #12 | #16 | #18 | #2 | #9 | #12 | #18 |
| $\Delta_q$ max | 24.5 | 43.5 | 18.8 | 17.8 | 33.2 | 38.1 | 3.83 | 15.9 | 34.5 | 35.3 | 4.25 | 9.4 | 9.9 | 11.7 | 18.2 | 9.2 |
| $\Delta_q$ min | 2.1 | 1.1 | 1.6 | 2.1 | 2.1 | 2.1 | 1.17 | 2.3 | 2.1 | 2 | 1.06 | 2 | 2 | 2 | 1.8 | 2.1 |
| $\Delta_q$ average | 18.7 | 11.6 | 4.4 | 11.9 | 11.4 | 9.6 | 1.62 | 9.1 | 8.1 | 10.2 | 1.71 | 6.2 | 3.8 | 4.2 | 5.2 | 5.1 |
| $\Delta_q$ Std dev | 4 | 11.8 | 3.4 | 3.9 | 8.6 | 9.3 | 0.42 | 3.3 | 5.8 | 7.2 | 0.53 | 1.9 | 2 | 1.9 | 3.2 | 1.9 |
| $\Delta_q$ nb / %déf | 15 | 18 | 10 | 15 | 13 | 13 | 27 | 15 | 14 | 10 | 15 | 10 | 4 | 6 | 4 | 9 |
| $\Delta_\varepsilon$ max | 0.134 | 0.265 | 2.024 | 0.23 | 0.363 | 0.218 | 0.162 | 0.183 | 0.223 | 0.232 | 0.313 | 0.348 | 0.673 | 0.472 | 0.619 | 0.28 |
| $\Delta_\varepsilon$ min | 0.04 | 0.003 | -0.005 | 0.02 | 0.014 | 0.011 | -0.012 | 0.019 | 0.02 | 0.027 | 0.011 | 0.024 | 0.026 | 0.066 | 0.105 | 0.028 |
| $\Delta_\varepsilon$ average | 0.065 | 0.054 | 0.1 | 0.069 | 0.079 | 0.078 | 0.037 | 0.064 | 0.071 | 0.103 | 0.067 | 0.105 | 0.265 | 0.153 | 0.219 | 0.105 |
| $\Delta_\varepsilon$ Std dev | 0.022 | 0.036 | 0.19 | 0.037 | 0.058 | 0.043 | 0.03 | 0.035 | 0.038 | 0.055 | 0.046 | 0.07 | 0.206 | 0.079 | 0.103 | 0.058 |
| $\Delta_\varepsilon$ nb / %déf | 15 | 18 | 10 | 15 | 13 | 13 | 27 | 15 | 14 | 10/1 | 15 | 10 | 4 | 6 | 4 | 9 |

| | $\sigma'_3 = 60$ kPa | | | |
|---|---|---|---|---|
| Velocity | 0.083 mm/min | | 0.18 mm/min | |
| size | H20, D10 | H8.4, D5 | H20, D10 | H8.4, D5 |
| # | #1 | #4 | #1 | #4 |
| $\Delta_q$ max | 34.5 | 60.1 | 32.4 | 38 |
| $\Delta_q$ min | 2.3 | 2 | 2 | 2.3 |
| $\Delta_q$ average | 14.6 | 11.4 | 10.1 | 9.6 |
| $\Delta_q$ Std dev | 10.5 | 9.9 | 8.4 | 7 |
| $\Delta_q$ nb / %déf | 16 | 23 | 21 | 19 |
| $\Delta_\varepsilon$ max | 0.21 | 0.195 | 0.112 | 0.167 |
| $\Delta_\varepsilon$ min | 0.004 | 0.007 | 0.007 | 0.018 |
| $\Delta_\varepsilon$ average | 0.062 | 0.043 | 0.047 | 0.054 |
| $\Delta_\varepsilon$ Std dev | 0.043 | 0.028 | 0.029 | 0.027 |
| $\Delta_\varepsilon$ nb / %déf | 16 | 23 | 21 | 19 |

**Table 2a: Results with stick-slip and d=630-840μm, at the top, $\sigma'_3 = 30$ kPa and at the bottom, $\sigma'_3 = 60$ kPa** : Results are classified as a function of the confining stress, the strain rate, and the n° of the test. They show statistical data (maximum amplitude, minimum amplitude, average, and standard deviation) of stress falls $\Delta q$ in kPa and the deformation between two falls, *i.e.* $\Delta \varepsilon$ in %. The size is measured in cm.

| | | | | | $\sigma'_3 = 30$ kPa | | | | | | | | |
|---|---|---|---|---|---|---|---|---|---|---|---|---|---|
| Velocity | 0.083 mm/min | | | | 0.18 mm/min | | | | | 0.50 mm/min | | | |
| Size | H20, D10 | H20, D10 | H20, D10 | H8.4, D5 | H20, D10 | H20, D10 | H20, D10 | H8.4, D5 | H20, D10, d3 | H20, D10 | H20, D10 | H20, D10 | H20, D10, d3 |
| # | #19 | #25 | #21 | #26 | #19 | #25 | #21 | #26 | #24 | #19 | #25 | #21 | #24 |
| $\Delta_q$ max | 0.52 | 0.53 | 2.2 | 2.95 | 1.13 | 0.78 | 2.6 | 3.8 | 19.9 | 0.6 | 0.45 | 4.29 | 19.3 |
| $\Delta_q$ min | 0.07 | 0.1 | 0.06 | 1.26 | 0.09 | 0.08 | 0.08 | 1.27 | 2.1 | 0.09 | 0.1 | 0.11 | 2.1 |
| $\Delta_q$ average | 0.16 | 0.19 | 0.25 | 1.73 | 0.18 | 0.18 | 0.52 | 1.74 | 11.5 | 0.24 | 0.22 | 1.28 | 9.4 |
| $\Delta_q$ Std dev | 0.07 | 0.08 | 0.19 | 0.45 | 0.06 | 0.09 | 0.41 | 0.48 | 3.5 | 0.12 | 0.09 | 0.85 | 3.6 |
| $\Delta_q$ nb / %déf | 110 | 71 | 94 | 27 | 45 | 44 | 60 | 13 | 20 | 20 | 17 | 21 | 21 |
| $\Delta_\varepsilon$ max | 0.028 | 0.109 | 0.054 | 0.424 | 0.07 | 0.089 | 0.091 | 0.381 | 0.098 | 0.191 | 0.145 | 0.099 | 0.109 |
| $\Delta_\varepsilon$ min | 0.001 | 0.001 | 0.001 | 0.004 | 0.007 | 0.006 | 0.001 | 0.017 | 0.013 | 0.025 | 0.026 | 0.022 | 0.023 |
| $\Delta_\varepsilon$ average | 0.008 | 0.014 | 0.011 | 0.037 | 0.022 | 0.023 | 0.017 | 0.075 | 0.049 | 0.051 | 0.06 | 0.048 | 0.048 |
| $\Delta_\varepsilon$ Std dev | 0.005 | 0.011 | 0.007 | 0.039 | 0.012 | 0.013 | 0.01 | 0.058 | 0.018 | 0.024 | 0.028 | 0.018 | 0.016 |
| $\Delta_\varepsilon$ nb / %déf | 110 | 71 | 94 | 27 | 45 | 44 | 60 | 13 | 20 | 20 | 17 | 21 | 21 |

**Table 2b : Results of experiments without stick-slip**: sand, glass beads with d =3 mm of diameter, *i.e.* material B and R. $\Delta q$ is in kPa and $\Delta \varepsilon$ in %.

## References

[1] K.H. Roscoe, A.N. Schofield, C.P. Wroth, Geotechnique 8, 22-53, (1958); P. Evesque, *Poudres & Grains NS1*, http://prunier.mss.ecp.fr /poudres&grains/ poudres-index.htm , and refs. there in.
[2] A. Duschesne, *Etude du Comportement Mécanique d'un Combustible Granulaire soumis à des Sollicitations d'origine Thermique dans un Propulseur Thermonucléaire Spatial* – PhD thesis, Ecole Centrale Paris, (1998).
[3] C. Thorton & L. Zhang, "A DEM comparison of different shear testing devices", in *Powders & Grains 2001*, Y. Kishino ed., Balkema, Rotterdam, 2001), pp. 177-180
[4] H. Tsunekawa & K. Iwashita, "Numerical simulations of triaxial test using two and three dimensional DEM", in *Powders & Grains 2001*, (Y. Kishino ed., Balkema, Rotterdam, 2001), pp. 183-190
[5] T.G. Sitharam, S.V. Dinesh & N. Shimizu, in *Powders & Grains 2001*, Y. Kishino ed., (Balkema, Rotterdam, 2001), pp. 241-245
[6] B.N.J. Persson, *Sliding Friction :Physical principles and Applications*, pp. 21, (Springer eds, Nano-Science and Technology, (1998).

The electronic arXiv.org version of this paper has been settled during a stay at the Kavli Institute of Theoretical Physics of the University of California at Santa Barbara (KITP-UCSB), in june 2005, supported in part by the National Science Fundation under Grant n° PHY99-07949.

*Poudres & Grains* can be found at :
http://www.mssmat.ecp.fr/rubrique.php3?id_rubrique=402